\documentstyle[12pt]{article}
\def\text#1{\hbox{#1}}
\def\subtext#1{\hbox{\scriptsize#1}}

\catcode`\@=11 
\def\maketitle{\par
 \begingroup
 \def\thefootnote{\fnsymbol{footnote}}
 \def\@makefnmark{\hbox
 to 0pt{$^{\@thefnmark}$\hss}}
 \if@twocolumn
 \twocolumn[\@maketitle]
 \else
 \global\@topnum\z@ \@maketitle \fi\thispagestyle{plain}\@thanks
 \endgroup
 \setcounter{footnote}{0}
 \let\maketitle\relax
 \let\@maketitle\relax
 \gdef\@thanks{}\gdef\@author{}\gdef\@title{}\let\thanks\relax}
\def\@maketitle{
 \null
 \vskip 2em \begin{center}
 {\LARGE \@title \par} \vskip 1.5em {\large \lineskip .5em
\begin{tabular}[t]{c}\@author
 \end{tabular}\par}
 \vskip 1em {\large \@date} \end{center}
 \par
 \vskip 1.5em}
\def\abstract{\if@twocolumn
\section*{Abstract}
\else \small
\begin{center}
{\bf Abstract\vspace{-.5em}\vspace{0pt}}
\end{center}
\quotation
\fi}
\def\endabstract{\if@twocolumn\else\endquotation\fi}

\mark{{}{}}

\newcount\@tempcntc
\def\@citex[#1]#2{\if@filesw\immediate\write\@auxout{\string\citation{#2}}\fi
  \@tempcnta\z@\@tempcntb\m@ne\def\@citea{}\@cite{\@for\@citeb:=#2\do
    {\@ifundefined
       {b@\@citeb}{\@citeo\@tempcntb\m@ne\@citea\def\@citea{,}{\bf ?}\@warning
       {Citation `\@citeb' on page \thepage \space undefined}}%
    {\setbox\z@\hbox{\global\@tempcntc0\csname b@\@citeb\endcsname\relax}%
     \ifnum\@tempcntc=\z@ \@citeo\@tempcntb\m@ne
       \@citea\def\@citea{,}\hbox{\csname b@\@citeb\endcsname}%
     \else
      \advance\@tempcntb\@ne
      \ifnum\@tempcntb=\@tempcntc
      \else\advance\@tempcntb\m@ne\@citeo
      \@tempcnta\@tempcntc\@tempcntb\@tempcntc\fi\fi}}\@citeo}{#1}}
\def\@citeo{\ifnum\@tempcnta>\@tempcntb\else\@citea\def\@citea{,}%
  \ifnum\@tempcnta=\@tempcntb\the\@tempcnta\else
   {\advance\@tempcnta\@ne\ifnum\@tempcnta=\@tempcntb \else \def\@citea{--}\fi
    \advance\@tempcnta\m@ne\the\@tempcnta\@citea\the\@tempcntb}\fi\fi}

\def\@cite#1#2{$^{#1\if@tempswa , #2\fi}$}

\catcode`\@=12 
\begin{document}

$$\phantom{a}$$
\vspace{-5cm}
\begin{flushright}
SLAC-PUB-6670\\
CERN-TH.7451/94\\
TAUP-2201-94\\
OSU-RN-293/94\\
hep-ph/9409376
\end{flushright}
\vspace{-1.0cm}

\author{John Ellis\thanks{%
Permanent address: Theory Division,
CERN, CH-1211, Geneva 23, Switzerland; \break
e-mail: johne@cernvm.cern.ch},
\ Marek Karliner\thanks{%
Permanent address: School of Physics and Astronomy,
Raymond and Beverly Sackler Faculty of Exact Sciences,
Tel Aviv University,
Ramat Aviv, Tel Aviv, Israel; e-mail: marek@vm.tau.ac.il}\
\ and Mark A. Samuel\thanks{%
Permanent address: Department of Physics, Oklahoma State University,
Stillwater, Oklahoma 74078, USA;
e-mail: physmas@mvs.ucc.okstate.edu}  \\ 
Stanford Linear Accelerator Center\\Stanford University, Stanford,
California 94309, USA
\medskip
\\and
\and Eric Steinfelds  \\ 
Department of Physics\\Oklahoma State University\\Stillwater,
Oklahoma 74078, USA}
\title{The Anomalous Magnetic Moments of the Electron and the
Muon  --  Improved QED Predictions using
Pad\'e Approximants\thanks{%
Work supported by the Department of Energy, contract DE-AC03-76SF00515.} }
\date{} %
\maketitle
\vspace{-0.3cm}
\begin{center}
{\large \bf Abstract}
\end{center}
We use Pad\'e Approximants to obtain improved predictions for the
anomalous magnetic moments of the electron and the muon. These are needed
because of the very precise experimental values presently obtained for
the electron, and soon to be obtained at BNL for the
muon. The Pad\'e prediction for the QED contribution to the
anomalous magnetic moment of the muon differs significantly from
the naive perturbative prediction.

\newpage\

Two of the most important tests of
quantum electrodynamics (QED) are the comparisons
between theory and experiment of the anomalous magnetic
moments of the electron and the muon, $a_e$
and $a_\mu $ respectively, where $a = (g{-}2)/2.$
The latest Penning trap measurements of the electron and
positron anomalies obtained by the University of
Washington group\cite{UofW} are:
\begin{equation}
\label{eq1a}a_{e^{-}}^{\subtext{expt}}=1159652188.4(4.3)\times 10^{-12}
\end{equation}
and
\begin{equation}
\label{eq2a}a_{e^{+}}^{\subtext{expt}}=1159652187.9(4.3)\times 10^{-12}
\end{equation}
The figures in brackets represent the error in the last 2 figures,
a convention we will follow throughout this paper. Taking
the average of eqs (\ref{eq1a}) and (\ref{eq2a}), one finds
\begin{equation}
\label{eq3}a_e^{\subtext{expt}}=1159652188.2(3.0)\times 10^{-12}
\end{equation}
The most accurate measurement for the muon anomaly comes from the CERN
$g{-}2$ experiment\cite{CERNgII}  in which it was found that
\begin{equation}
a_{\mu ^{-}}^{\subtext{expt}}=1165936(12)\times 10^{-9}
\end{equation}
and
\begin{equation}
a_{\mu ^{+}}^{\subtext{expt}}=1165910(11)\times 10^{-9}
\end{equation}
and the combined result is
\begin{equation}
a_\mu ^{\subtext{expt}}=1165923(9)\times 10^{-9}
\end{equation}
where correlations are taken into account in combining the errors.
A new $g{-}2$ muon experiment is being done
at Brookhaven National Laboratory (BNL)\cite{BNL},
and an improvement in the accuracy by a factor of about 20 is
expected. In order to compare properly theory and experiment, one must
improve correspondingly the accuracy of the theoretical predictions.

\par
In an
heroic feat, Kinoshita\cite{Kinoshita}
 has calculated $a_e$ in eighth order and
Kinoshita, Nizic, Okamoto\cite{KNO}
and Marciano\cite{KinoshitaMarciano} have calculated $a_\mu $ in
eighth order. Moreover, there have been some recent improvements in the
analytic calculations\cite{improvedaI,improvedaII}
of $a_e$ and $a_\mu $.

\par
There have recently been several papers estimating coefficients in
Perturbative Quantum Field Theory (PQFT) using
Pad\'e Approximants\cite{PadeInQFT,markop}.
This procedure is known to give
significant improvements on naive perturbative calculations in
many condensed-matter applications\cite{condm},
removes a large part of the
discrepancy between experiment and QED calculations of the
ortho-positronium decay rate\cite{markop,KriplovichMilstein}
and agrees well with other estimates of
higher-order perturbative coefficients
in QCD\cite{PadeInQFT,EstimCorr}.

\par
In
this paper we will use Pad\'e Approximants (PA's) to estimate, not
just the next
term in the perturbation series, but the entire sum of the series
(as is frequently done in condensed-matter applications), for both $%
a_e$ and $a_\mu$. We obtain in this way a more accurate theoretical
prediction of the QED contribution to $a_{\mu}$, in particular,
which lies outside the errors quoted previously.

\par
The first step is to obtain an accurate value for the fine-structure
constant $\alpha $. The two most precise measurements of $\alpha $
are\cite{alfaEXPa}
\begin{equation}
\label{eq4}\alpha ^{-1}=137.0359979(32)
\end{equation}
and\cite{alfaEXPb}
\begin{equation}
\label{eq5}\alpha ^{-1}=137.0359840(50)
\end{equation}
We note that these two values differ by more than 2 standard
deviations, but nevertheless
take the average of eqs (\ref{eq4}) and (\ref{eq5}) to obtain
\begin{equation}
\label{eq7}\alpha _{\subtext{exp}}^{-1}=137.0359939(27)
\end{equation}
The accuracy of this result limits the precision of tests of QED in the
case of $a_e$, where both theory and experiment are
extremely precise. The perturbation series for $a_e$
is\cite{Kinoshita}
\begin{equation}
a_e=\frac 12(\frac \alpha \pi )-0.328478965(\frac \alpha \pi
)^2+1.17611(42)(\frac \alpha \pi )^3-1.434(138)(\frac \alpha \pi )^4
\end{equation}
and the error in the theoretical prediction is dominated by the
error in $\alpha_{\subtext{exp}}$.

\par
The [N/M] Pad\'e Approximant to a series
\begin{equation}
S=S_0+S_1x+S_2x^2+...+S_{N+M}x^{N+M}
\end{equation}
is given by
\begin{equation}
[N/M]=\frac{a_0+a_1x+...+a_Nx^N}{1+b_1x+...+b_Mx^M}
\end{equation}
where one chooses the coefficients $a_i$, $b_j$ so that
\begin{equation}
[N/M]=S+O(x^{N+M+1})
\end{equation}
One can use such a PA either to predict the next coefficient
$S_{N+M+1}$ or to evaluate $[N/M]$ for
the relevant value of $x$ (in our case $x=\frac \alpha \pi $ ),
and obtain an
estimate for the sum of the series. Here we do the
latter. The PA's are known to accelerate the
convergence of many series by including the effects of higher
(unknown) terms, thus providing a more accurate estimate
of the series\cite{condm}.
The PA's also provide reliable estimates of many
asymptotic series, as is the case in QED\cite{markop}
 and QCD\cite{PadeInQFT}.

\par
For our application, we first construct PA's to $a_e$ after
removing an overall multiplicative factor of
$({\alpha\over\pi})$. Our result for the [1/2] PA is
\begin{equation}
[1/2]=1159652169.1(24.0)\times 10^{-12}
\end{equation}
and the [2/1] PA agrees very well with the [1/2]:
\begin{equation}
[2/1]=1159652169.0(24.0)\times 10^{-12}
\end{equation}
The errors consist of 22.8 from $\alpha $ and 7.4 from the theoretical
uncertainty. To obtain $a_e$ one must add the contribution due to
muon diagrams
\begin{equation}
\Delta a_e(\text{muon})=2.8\times 10^{-12}
\end{equation}
the contribution due to $\tau$ diagrams
\begin{equation}
\Delta a_e(\text{tau})=0.01\times 10^{-12}
\end{equation}
the hadronic contribution
\begin{equation}
\Delta a_e(\text{hadron})=1.6(2)\times 10^{-12}
\end{equation}
and the purely weak contribution
\begin{equation}
\Delta a_e(\text{weak})=0.05\times 10^{-12}
\end{equation}
for a total of
\begin{equation}
\label{eq8}\Delta a_e=4.5(2)\times 10^{-12}
\end{equation}
Thus the theoretical prediction for $a_e$ is
\begin{equation}
\label{eq1}a_e=1159652173.5(24.0)\times 10^{-12}
\end{equation}
Comparing eq (\ref{eq1}) with eq (\ref{eq3}),
 we see that there is beautiful
agreement between theory and experiment:
\begin{equation}
\label{eq2}a_e^{\subtext{expt}}-a_e=14.7(24.0)\times 10^{-12}\text{ }%
(0.61\sigma )
\end{equation}

\par
As noted before, the error in eq (\ref{eq1})
is dominated by the error in $\alpha $.
If one now assumes
that QED is correct, and hence that theory and experiment
agree, one obtains a new and
more accurate value of $\alpha $: $\alpha _{\subtext{%
th}}$, where the [1/2] PA gives
\begin{equation}
\label{eq9}\alpha _{\subtext{th}}=137.03599228(86)
\end{equation}
and the [2/1] PA leads to
\begin{equation}
\label{eq6}\alpha _{\subtext{th}}=137.03599227(86)
\end{equation}
Comparing eq (\ref{eq9}) with eq (\ref{eq7}),
 one sees that there is beautiful
agreement with the less-precise experimental value.
\begin{equation}
\alpha_{\subtext{expt}}^{-1}-\alpha_{\subtext{th}}^{-1}=
16(28)\times 10^{-7}%
\text{ (0.57}\sigma )
\end{equation}
corresponding to the good agreement in eq (\ref{eq2}). We note in
passing that this provides an {\em a posteriori} justification for
averaging naively the two most accurate
measurements\cite{alfaEXPa,alfaEXPb}
of $\alpha$, and that
the difference between the values of
$\alpha_{\subtext{th}}^{-1}$ extracted using the perturbative
series and the PA's is just $3 \times 10^{-8}$.

\par
We now turn to the anomalous magnetic moment of the
muon, $a_\mu $. As is usual, we first consider the
difference\cite{KinoshitaMarciano,improvedaII}
\begin{equation}
\label{eq10}a_\mu -a_e=1.09433583(7)(\frac \alpha \pi )^2+22.869265(4)(\frac
\alpha \pi )^3+127.00(41)(\frac \alpha \pi )^4
\end{equation}
In constructing a PA to this series, we must first remove an
overall factor $({\alpha\over\pi})^2$ from the
perturbative series. In this way, we obtain the
[1/1] PA value
\begin{equation}
(a_{\mu} - a_e)
[1/1]=6194839(12)\times 10^{-12}
\end{equation}
whereas the value from the series in eq (\ref{eq10}) is
\begin{equation}
a_\mu -a_e=6194791(12)\times 10^{-12}
\end{equation}
Adding $a_e$ from eq (\ref{eq1}), after
subtracting $\Delta a_e$ from eq (\ref{eq8}), we obtain
\begin{equation}
a_\mu ^{\subtext{QED}}=1165847008(12)(27)\times 10^{-12}
\end{equation}
where the first error is due to numerical integrations used in
evaluating the perturbative series, and the second error is
due to $\alpha$. This should be compared
the value of Kinoshita and Marciano\cite{KinoshitaMarciano}
\begin{equation}
a_\mu ^{\subtext{QED}}=1165846955(44)(27)\times 10^{-12}
\end{equation}
We note that the difference between these two estimates of
$a_{\mu}^{\subtext{QED}}$ is considerably larger than the error
propagated from $\alpha$. The reason for our smaller error
is that we have used the new more precise values in ref.~[8].

\par
If one now adds the hadronic\cite{HadronicCont}
 and the weak\cite{KinoshitaMarciano}
 contributions
\begin{equation}
\Delta a_\mu (\text{had})=7011(76)\times 10^{-11}
\end{equation}
and
\begin{equation}
\Delta a_\mu (\text{weak})=195(10)\times 10^{-11}
\end{equation}
one obtains the theoretical value
\begin{equation}
a_\mu =116591907(77)\times 10^{-11}
\end{equation}
The error is dominated by the error in $\Delta a_\mu ($had$)$, and
new, more
precise experiments are underway in Novosibirsk and Frascati\cite{novos}
to reduce this error.
Comparing eq (33) with eq (6), we obtain
\begin{equation}
a_\mu^{\subtext{expt}}-a_\mu=4(9)\times 10^{-9}\text{ }%
(0.4\sigma )
\end{equation}
The error in the difference between theory and experiment is dominated
by the experimental error in eq (6), which should be reduced by a
factor of 20 in the forthcoming BNL experiment\cite{BNL}.

\par
In summary, we have used PA to obtain new more precise values for the QED
values of $a_e$ and $a_\mu $ . These PA values, in effect, estimate the
unknown higher-order contributions, and should be more precise than
the naive perturbative values used previously. It would be interesting
to compare our estimates with values obtained in a different way,
for example using the effective charge approach\cite{EstimCorr}
 which agrees
very well with PA's in QCD applications\cite{PadeInQFT}. Although
smaller than some
of the other uncertainties, the shift we find in $a_{\mu}$, in
particular, is significantly larger than other theoretical
uncertainties and the error due to $\alpha$.

J. E., M. K. and M. S. would like to thank the theory
group at SLAC for its
very kind hospitality.
The  research of M.K. was supported in part
by grant No.~90-00342 from the United States-Israel
Binational Science Foundation (BSF), Jerusalem, Israel,
and by the Basic Research Foundation administered by the
Israel Academy of Sciences and Humanities.
The research of  M.A.S. was supported by the U.S. Department of
Energy under Grant No. DE-FG02-94ER40852.

\end{document}